# Exploiting solute segregation and partitioning to the deformation-induced planar defects and nano-martensite in designing ultra-strong Co-Ni base alloys

Akshat Godha[1*], Mayank Pratap Singh[1], Karthick Sundar[1] Shashwat Kumar Mishra[1], Praveen Kumar[1], Govind B[2], Surendra Kumar Makineni[1*]

[1]Department of Materials Engineering, Indian Institute of Science Bangalore – 560012

[2]Materials and Metallurgy Group, Vikram Sarabhai Space Centre, Thiruvananthapuram – 695022

**\*Corresponding Authors:** akshatgodha@iisc.ac.in; skmakineni@iisc.ac.in

**Abstract**

Stronger metallic alloys are always sought as structural materials to enable efficient, safe, and sustainable operations across various aerospace, automotive, energy, and defence sectors. Among these single-phase, multi-elements (three or more) with high concentrations show exceptional tensile strength up to ~ 0.8-1.2 GPa. However, they possess a very low 0.2% yield strength (YS), i.e., they can be permanently deformed at very low-stress levels of ~ 300 to 600 MPa. Here, we reveal by exploiting atomic-scale solute interactions with the deformation-induced structures to design ultra-strong single-phase alloys with YS > 2 GPa. This was achieved by controlled thermomechanical processing that introduces stacking-faults (SFs), nano-twins (NTs), and nano-martensite ε-laths (NMLs) during cold deformation followed by facilitating solute segregation/partitioning to them by tempering at intermediate temperature. We demonstrate the phenomena in a low stacking faulty energy multi-component (face-centered-cubic, fcc structured) Co-33Ni-24Cr alloy (all in at.%) containing 5at.% Mo as a solute. It is also shown that the degree of strengthening after tempering scales up with the fraction of these structures (before tempering) in the alloy microstructure that can be tuned by the amount and temperature of cold deformation. Cold-rolling with 45% and 65% thickness reduction, followed by tempering at 600°C for 4 hours, led to an YS of ~ 1.5 GPa and ~ 2 GPa with elongation to fracture (%El) ~ 14% and ~7%, respectively. The YS is further enhanced to ~ 2.2 GPa without reduction in %El upon cryo-rolling followed by tempering. The alloy microstructure is stable at 600°C up to 100 hours and also retains an YS of ~ 1.5 GPa with %El of ~ 18% during tensile test at 600°C. The derived high YS and high-temperature stability are critically a consequence of solute partitioning to the NMLs that we termed as Solute-Partitioned NMLs (SP-NMLs) in the microstructure. Based on the exploration and revelation of structure-dependent segregation/partitioning of deformation-induced planar defects and NMLs in the alloy, we propose that this strategy would not be limited to the present alloy composition and can be effective in other low SFE alloys with suitable solutes that can open avenues for designing ultra-strong materials.

## 1. Introduction

For structural applications, a major section of high-strength alloys, such as steels, superalloys, lightweight alloys, etc., has a base element of iron, nickel, cobalt, titanium, aluminium, or their combinations. Increasing the strength of the alloy largely relies on imparting resistance to the motion of dislocations in their microstructure during external load [1]. The most practised strengthening strategies comprise solid-solution hardening, grain-boundary strengthening, precipitation/dispersion hardening, and mechanical working, including severe plastic deformation (SPD) that induces a high density of dislocations and grain refinement [2–5]. Controlled post-heat cycles to the mechanically deformed alloys also allow heterogeneous microstructures with gradients in the distribution of grain sizes and the formation of harder/softer zones [6,7]. Tuning of the volume fraction of these features provides ways to balance the strength-ductility trade-off of the alloy. Additionally, twinning and martensitic transformations during deformation are most exploited for alloys with low stacking fault energies (10 to 30 mJ/m$^2$) [8]. Both transformations have also been shown as excellent modes of plastic strain accumulation during tensile loading, leading to a high degree of strain hardening and strength. These were coined as twinning-induced plasticity (TWIP) [9] and transformation-induced plasticity (TRIP) effects [10]. Based on these, several remarkable alloys were developed, such as advanced-high-strength-steels (AHSS) [9–11] and multi-principle-element (MPE) alloys [12–20] with ultimate tensile strength (UTS) values of ~ 0.8-1.2 GPa. However, these alloys start yielding at very low-stress levels (0.2% yield stress (YS) ~ 300 to 600 MPa), i.e., low elastic regime, indicating their easy tendency towards getting permanently (plastically) deformed [9].

During the last decade, several developments related to thermo-mechanical processing and/or tuning alloy compositions were demonstrated to increase their YS [21–28]. For example, generating nano-twin bundles by dynamic plastic deformation, inducing gradient nano-twinned structures along the radial direction by torsion, etc., can increase the yield stress up to ~1 GPa [21]. Based on alloy compositions, another way to get stronger alloys is by introducing martensite either thermally by quenching or deformation, resulting in highly dislocated martensite in the microstructure. In the former case, subsequent heat treatment leads to a high density of nano-precipitates [24–26], while in the latter, dislocation hardening increases the stress barrier to the movement of dislocations, leading to yield-stress values beyond 2 GPa [27].

Here, we show that facilitating atomic-scale solute segregation/partitioning into the deformation-induced planer defects and nano-martensite-laths (NMLs) produces ultra-strong alloys with YS values > 2GPa. A low SFE alloy of composition Co-33Ni-24Cr-5Mo (at.%) was chosen for the demonstration, where Mo is a solute. The composition is similar to the commercially available trademark MP35N$^{TM}$ alloy. Although this alloy was developed five decades ago and several works were published, there is no consensus about the operative strengthening mechanisms. One common finding among many reports is that there are no precipitates in the tempered/annealed microstructure and, hence, no contribution from the strengthening by precipitates. Several mechanisms were proposed, such as Suzuki strengthening by segregation of solutes to the deformation-induced SFs, but they could not be verified explicitly. For example, Singh et al. [29] and Asgari et al. [30] hypothesized that segregation of Cr/Mo at the SFs could be responsible for the increased hardness on tempering, but no evidence was provided. Recently, Chiba's group [31] showed segregation evidence of Mo/Co at an SF, which was formed after 10% tensile deformation and annealing, using energy-dispersive X-ray-spectroscopy (EDS) in a Scanning-TEM, but couldn't show the segregation behavior at the SFs in the alloys that are cold-deformed and followed by annealing. However, they performed extensive simulations to rationalize the effect of Mo/Co segregation to the SFs on the hardening mechanism in the tempered alloy [32]. In contradiction, more recent works by several groups conducted the atomic-scale compositional analysis using an atom probe for the deformed/tempered alloy. They claimed the hardening is attributed to the segregation of Cr at the SFs/twins [33–35] and the segregation of Mo at the grain boundaries (GBs) rather than at the SFs [36].

In this work, contrary to the above illustrations and speculations, we explicitly reveal the formation of deformation-induced atomic-scale structures with their compositions in the microstructure that are responsible for the ultra-high YS of the alloy. We also show that the fraction of these structures is controlled by the temperature and degree of cold deformation, providing scope for designing alloys with desired YS values for structural applications.

## 2. Experimental Details
### 2.1 Alloy preparation, heat treatment, and processing

The alloy ingot with nominal composition Co38-Ni33-Cr24-Mo5 (at%) was prepared by vacuum induction melting (VIM) followed by vacuum arc remelting (VAR) at the Indian Space Research Organization (ISRO). The ingot weighing 1kg was melted multiple times to ensure the homogeneity of the composition. Following casting, the ingot was hot rolled at 1100°C followed by homogenized at 1200° for 20 h under vacuum. Small rectangular slabs were cut from the homogenized sample in an electric discharge machine (EDM). These slabs were then subjected to cold rolling (CR) up to 45% (hereafter referred to as 45CR) and 65% reduction (hereafter referred to as 65CR), followed by annealing at 600°C (hereafter referred to as 45CRT and 65 CRT for 4 hours annealed condition) up to 100h.

### 2.2 Mechanical properties

The hardness of the rolled and aged samples was measured using a Vickers microhardness tester with a load of 0.5kg. For the calculation of 0.2% yield strength (YS), room temperature tensile tests were conducted using Instron 5967 UTM (universal testing machine) at a constant crosshead displacement rate of $10^{-3}$ s$^{-1}$. Flat dog-bone-shaped tensile samples with a gauge length of 6mm, width of 2mm, and thickness of 1mm were cut in EDM such that the longitudinal axis of the samples is parallel to the rolling direction. Digital Image Correlation (DIC) was employed for precise strain measurements during the tensile tests. The gauge section of the samples was coated with a speckle pattern, and images were processed using VIC-2D software to perform full-field strain analysis. A subset size of 160 × 160 μm² with a step size of 7 pixels was selected, providing a spatial resolution of approximately 70–75 μm. The true strain rate calculated from DIC analysis was approximately $0.9 \times 10^{-3}$ s$^{-1}$, closely matching the crosshead displacement rate set for these tests.

## 2.3 Microstructural Characterization

### 2.3.1 X-ray Diffraction

Identification and evolution of the phase after rolling and aging were investigated using a Rigaku Smartlab x-ray diffractometer with Cu-K$_\alpha$ radiation equipped with Johansson optics to eliminate the contribution from K$_{\alpha2}$ and K$_\beta$ components.

### 2.3.2 Electron Channeling Contrast Imaging (ECCI)

After each processing step, the microstructure was characterized using a Carl Zeiss Gemini scanning electron microscope (SEM) equipped with a field emission gun (FEG) Source. Grain size and orientation were determined by electron back-scattered diffraction (EBSD) with a step size of 0.1μm. Electron channeling contrast imaging (ECCI) was employed to characterize the deformed structure over the large area and to prepare the site-specific samples from the defects regions for correlative transmission electron-microscopy (TEM) and atom probe tomography (APT) experiments. The ECCI was carried out at an operating voltage of 20kV, a probe current of 4 nA, and a working distance of 7mm. Details of the working procedure for ECCI are provided elsewhere [37]. For SEM and ECCI analysis, samples were prepared by mechanical grinding with Sic grid papers (up to 4000 mesh size) followed by cloth-polishing in alumina suspension of particle size 3 to 0.05um.

### 2.3.2 Transmission Electron Microscopy (TEM)

Diffraction contrast imaging was carried out in TEM using a T20 Tecnai (ThermoFisher make) instrument operated at 200kV. TEM specimens were prepared by sectioning 3mm disc samples via EDM, followed by mechanical polishing to a thickness of 100 μm. Final thinning for electron transparency is achieved using a twinjet electropolishing unit with an A2 solution composed of ethanol and perchloric acid solution. High-resolution imaging was carried out in an aberration probe-corrected TEM (Titan Themis, ThermoFisher make) in scanning-TEM (STEM) mode operated at 300kV. The drift-compensated images were acquired using Velox software with a high angle-annular-dark-field (HAADF) detector at a camera length of 160 mm.

### 2.3.4 Correlative TEM and Atom Probe Tomography (APT)

For the correlative TEM/APT experiment, site-specific APT needles were fabricated using a dual-beam SEM/focused-ion-beam (FIB) instrument (Scios Nanolab, ThermoFisher make). The cut lamella containing ROI was attached to the electropolished half-Mo grids, which were mounted in a specially designed correlative holder in-house. Subsequent sharpening of the lamella was accomplished by using Ga$^+$ ions accelerated at 30kV with beam currents ranging from 0.3 nA to 50 pA. Final surface cleaning was done at 2kV to minimize the Ga$^+$ damage regions. APT data were acquired using a LEAP 5000XR atom probe (Cameca Instruments Inc) operated in laser pulsing mode. A laser pulse repetition rate of 125kHz and pulse energy of 55pJ was used. The base temperature of the needles was kept at 80K, and the target detection rate was set to 0.5%. Data processing and reconstruction were carried out using the IVAS™ 3.8.10 software package.

## 3. Results

### 3.1 Microstructure of homogenized sample

Figure 1(a) shows a back-scattered electron (BSE) micrograph of the homogenized sample (as-received), revealing an equiaxed grain structure with an average size of ~40 μm. Figure 1(b) shows the EBSD grain orientation map with an overlaid image quality (IQ) map. Both SEM and EBSD confirm a single-phase microstructure throughout the sample, which is further supported by the X-ray diffraction pattern in Supplementary Figure S1 (a), where all characteristic indexed diffracted peaks correspond to the face-centered-cubic (fcc) phase.

### 3.2 Mechanical properties after cold rolling (CR) and tempering (CRT)

The alloy was subjected to cold rolling with 45% (45CR) and 65% (65CR) thickness reductions. Figure 1(c) illustrates the hardness evolution in 45CR and 65CR alloys during tempering at 600°C over different durations. The initial hardness of the homogenized sample (~180Hv) increases to ~440Hv for the 45CR and ~480Hv for the 65CR samples, which can be attributed to Taylor hardening [38], which results from increased defect density during cold rolling. Upon tempering at 600°C, the hardness increases to a peak value of ~540Hv for 45CR and ~590Hv for 65CR samples after 4h. These samples will henceforth be referred to as CRT (cold-rolled and tempered) samples. Furthermore, the hardness of the 65CRT retains up to ~580 Hv even after 100 hours of tempering, whereas the hardness of the 45CRT drops after 4 hours and up to ~470 Hv after 100 hours of tempering. Figure 1(d-f) show engineering stress-strain curves of 45/65 CR/CRT samples with respect to the homogenized sample. Note that only a small portion of the tensile curve is shown for the homogenized sample (%El is ~ 80%). The tensile YS of the homogenized sample increases from ~ 350MPa to 1.3 GPa and 1.5 GPa, respectively, after 45% and 65% CR. The YS increases by ~ 15% for 45CR after tempering (1.34 ± 0.046 GPa to 1.54 ± 0.026 GPa, higher by ~0.25 GPa) while the YS increases by ~ 30% for 65CR after tempering (1.53 ± 0.028 GPa to 1.98 ± 0.012 GPa, higher by ~0.45 GPa). This large difference in the YS increment between 45CRT and 65CRT is due to the difference in the type of defect structures formed during cold rolling as discussed below.

### 3.3 Microstructures after cold rolling (CR)

Supplementary Figure S1 (b-c) shows low magnification SEM micrographs of 45CR and 65CR samples showing similar microstructure of bright and thick shear bands. The electron-back-scattered-diffraction (EBSD) grain orientation maps, Supplementary Figure S1 (d-e), show a higher fraction of strained regions in 65CR as compared to 45CR. Figure 2(a-b) shows the defect structures in 45CR alloy. Bright-field images taken along the [110] zone axis reveal several planar faults and nano-twins (NTs) in edge-on conditions that are identified by streaks along 111 diffraction spots and additional twinning spots in the respective [110] diffraction patterns (see insets). Figure 2(a) highlights multiple Lomar-Cottrell (L-C) locks (indicated by white arrow) that form when two non-coplanar SFs on distinct {111} planes intersect with each other. An atomic-resolution high-angle-annular-diffraction (HAADF) Scanning-TEM (STEM) image centered on an L-C lock with two intersecting intrinsic SFs is also shown. Figure 2(b) shows a similar bright-field and HAADF STEM image centered on an NT, which extends from 13 layers (left) to 18 layers (right).

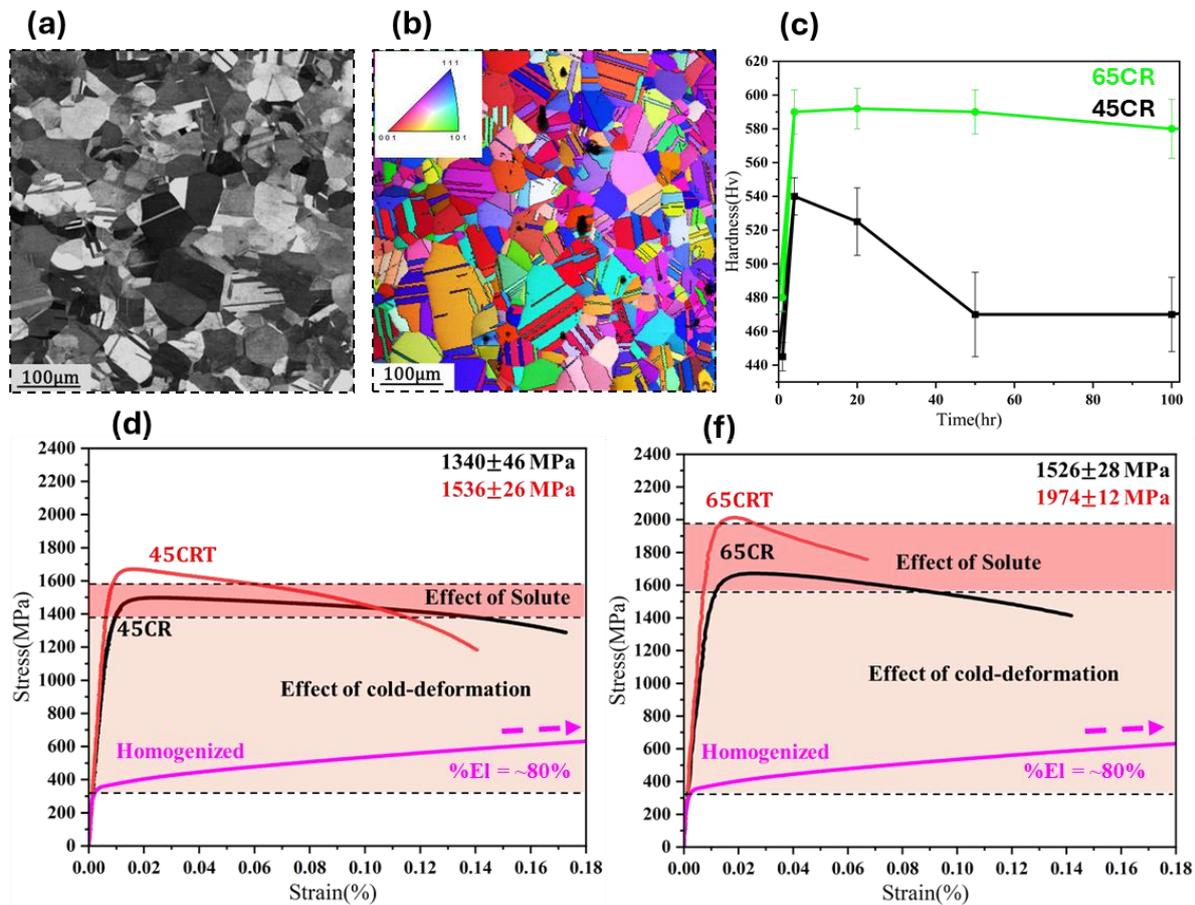

*Figure 1*: Ni-Co-Cr-Mo microstructure and mechanical properties after cold rolling and tempering: (a) Back-Scattered-Electron (BSE) SEM image of the homogenised No-Co-Cr-Mo alloy showing coarse-grained microstructure. (b) An electron-back-scattered-diffraction (EBSD) orientation map of the alloy reveals a single phase indexed as a face-centered-cubic (fcc) microstructure. (c) Evolution of hardness during tempering at 600°C after cold rolling with 45% (45CR) and 65% (65CR) thickness reductions. Engineering tensile stress-strain curves for (d) 45CR, 45CRT (tempered for 4 hours) alloys and (e) 65CR and 65CRT (tempered for 4 hours)

Similar to the 45CR alloy, several isolated SFs and NTs were observed in the 65CR alloy, Figure 2(c-d). Here, two-layer SFs of an extrinsic nature (ESF) were also identified. These ESFs can also act as embryos for forming NTs during deformation. In addition to the SFs and NTs, nano-martensite laths (NMLs) were also located across the microstructure in the 65CR. These are either seen in isolation or adjacent to NTs (NT/NML laminated structure), as shown in Figure 2(e-f). The atomic structure of ε-lath is hexagonally-close-packed (hcp) having the sequence ABABABAB…., different from the twinned region, which is fcc. The corresponding defect structures in 45Cr and 65CR alloys were retained after tempering at 600°C for 4 hours (see Figure S2).

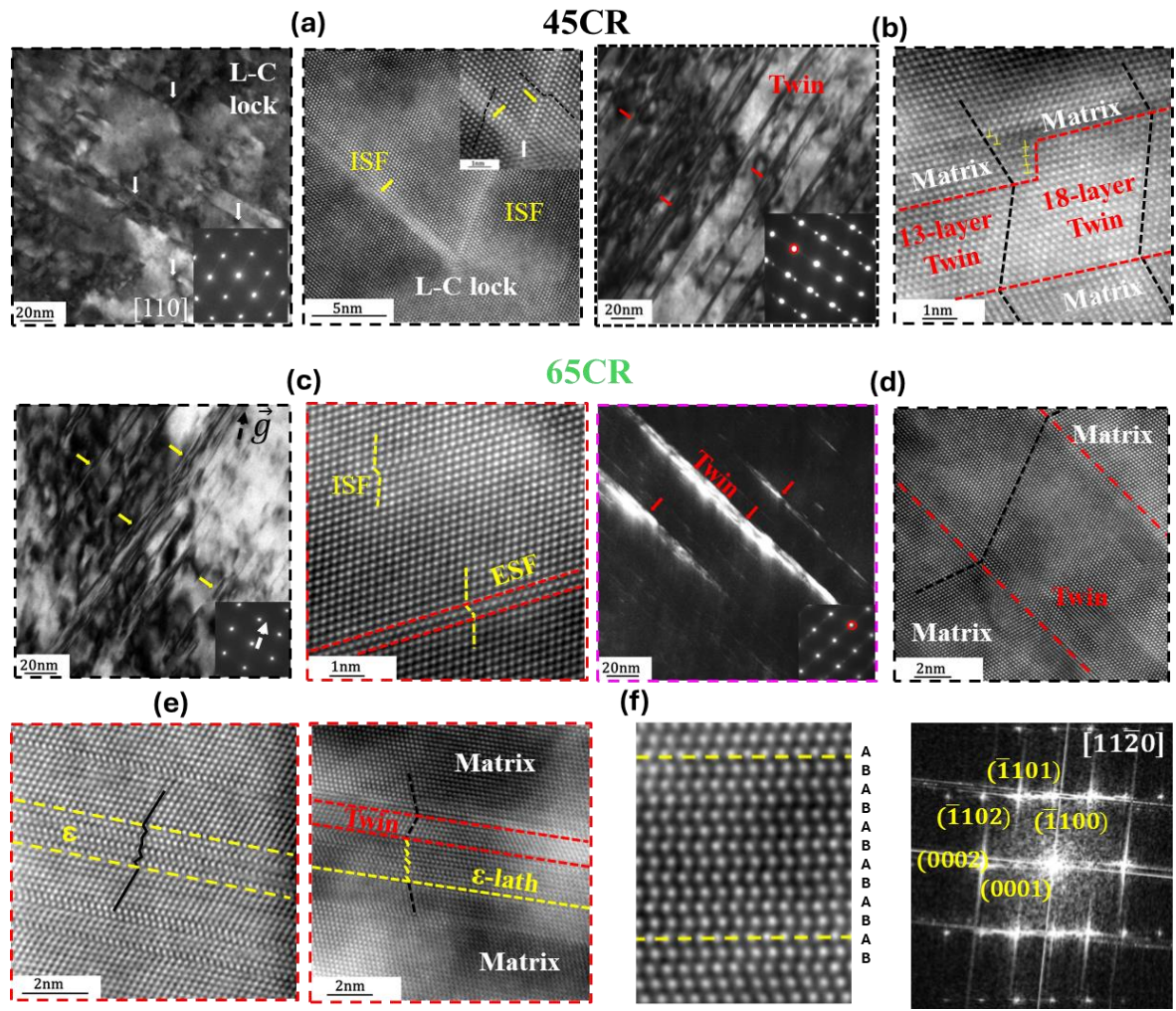

*Figure 2*: Microstructure after cold rolling (45% and 65% thickness reductions): Bright-field (BF) TEM and high-resolution HAADF STEM images for the 45CR alloy showing high number density of (a) stacking-faults (SFs) and Lomer-Cotrell (LC) locks, (b) nano-twins (NTs). Similarly, BF TEM and HAADF STEM Images for the 65CR alloy showing (c) stacking-faults (SFs) with both intrinsic and extrinsic nature, (d) nano-twins (NTs) and (e-f) hexagonally-close-packed (hcp) ε nano-martensite-laths (NMLs) formed in isolation or adjacent to NTs (NT/NML laminated structures).

## 3.4 Atomic-scale solute segregation/partitioning to the deformation-induced structures after tempering (CRT)

Correlative structural and compositional analysis of the deformation-induced structures at the atomic-scale was performed for the cold-rolled, followed by tempered alloys (CRT). The following section shows the structure and compositional analysis for SFs, NTs, and NMLs in 65CRT and SFs and NTs in 45CRT alloys.

### 3.4.1 Stacking Faults (SFs) and Nano-Twins (NTs)

Figure 3(a) shows a HAADF-STEM micrograph of a needle specimen along the [110] zone axis from the 65CRT alloy. A region marked by a black dashed square shows a bright contrast linear feature corresponding to an SF. The atomic structure across the same region reveals it as an ISF. The same needle was field evaporated in an atom probe. Figure 3(b) shows the reconstruction with the distribution of Co, Ni, Cr, and Mo atoms. A planar feature is identified with a 7.2 at.% iso-composition value of Mo. The compositional profile across the fault reveals a depletion of Cr/Ni by ~2at.%/1at.% and enrichment of Mo/Co by 1.8at.%/0.5at.% at the fault plane with respect to the surrounding matrix, respectively (other 1D and 2D elemental plots and are shown in supplementary Figure S3). The 2-D compositional maps of Mo/Cr taken in edge-on conditions centered on the SF show the depletion (Cr) and the enrichment (Mo) across the fault plane. While the surrounding region of the fault shows enrichment of Cr and depletion of Mo, indicating the occurrence of cross-diffusion of solutes to the SF plane from the surrounding matrix.

Another needle sample was taken from the twinned region of 65CRT alloy for the correlative analysis, Figure 3(c). A dark-field image taken from the twin diffraction spot along the [110] zone axis highlights the twin regions crossing the needle. From the bright contrast in the DF image, it appears the presence of two NTs (T1 and T2). The same needle was field evaporated, and Figure 3(d) shows the reconstruction with the distribution of Co, Cr, Ni, and Mo atoms. Both T1 and T2 were identified by two iso-compositional surfaces of 22at.% (cyan) and 19at.% (magenta) Cr, respectively. From the 1-D composition profile across the twinned regions, we observed two peaks of enrichment of Co/Mo and depletion of Cr/Ni. The first peak shows Mo increases up to 12at.% while the second one has Mo up to 8.5 at.%. A similar analysis of SFs and NTs in the 45CRT sample indicates Mo/Co segregation and Cr/Ni depletion at the fault planes, Supplementary Figure S4.

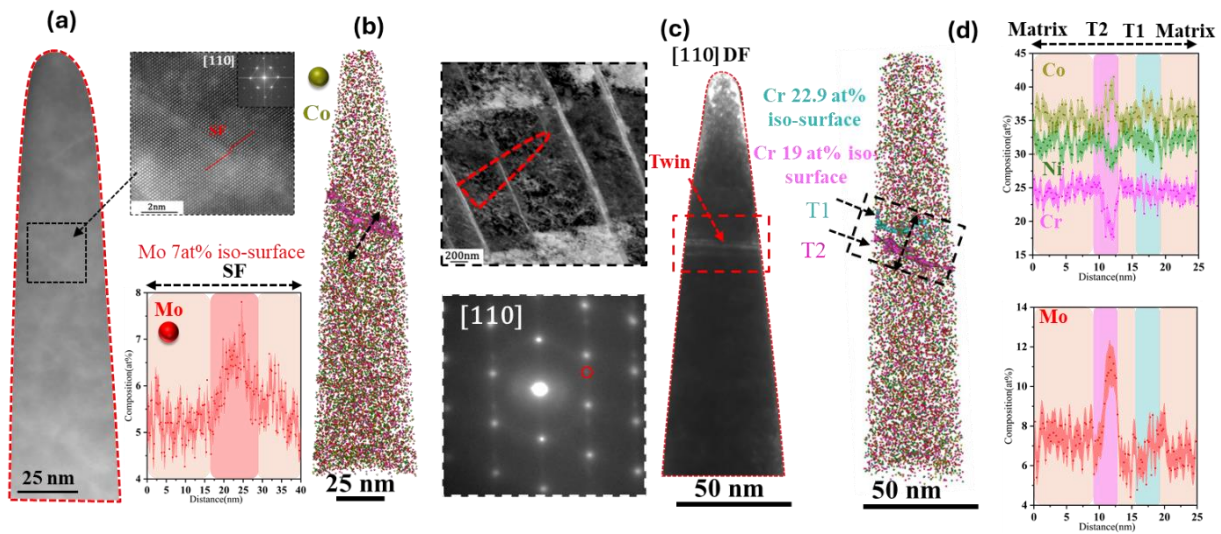

*Figure 3:* (a) An HAADF-STEM image of a needle specimen from a deformed region of the 65CRT alloy showing an SF in the edge-on condition. The atomic resolution HAADF STEM image taken from the same region shows that the SF is extrinsic in nature. (b) Atom probe reconstruction of the same needle after filed evaporation with the distribution of Co, Cr, Ni, and Mo atoms and an iso-compositional delineated by 7.2at.% of Mo corresponds to the extrinsic SF. The compositional profile across the SF shows Mo/Co enrichment at the fault plane while Cr/Ni depletes. (c-d) Similar correlative structural and compositional analysis on the Nano-Twins (NTs) from the 65CRT alloy reveals Mo/Cr enrichment at the NTs while Cr/Ni depletes.

### 3.4.2 Nano-Martensite-laths (NMLs)

Similarly, Figure 4(a) shows a low-magnification HAADF-STEM image taken from a deformed region acquired along the [110] zone axis of the matrix. The top part of the needle was identified as NML having the hcp atomic structure of ABAB…type shown as HR-STEM image (Figure 4(b)) from the region along with the corresponding FFT pattern indexed as [11$\bar{2}$0] zone axis. Below the hcp phase, we also found the presence of a laminated NT/NML structure. Before field evaporation of the sample needle, FIB milling followed by cleaning was carried out to make sure both NML and NT/NML structures could be captured in the atom probe.

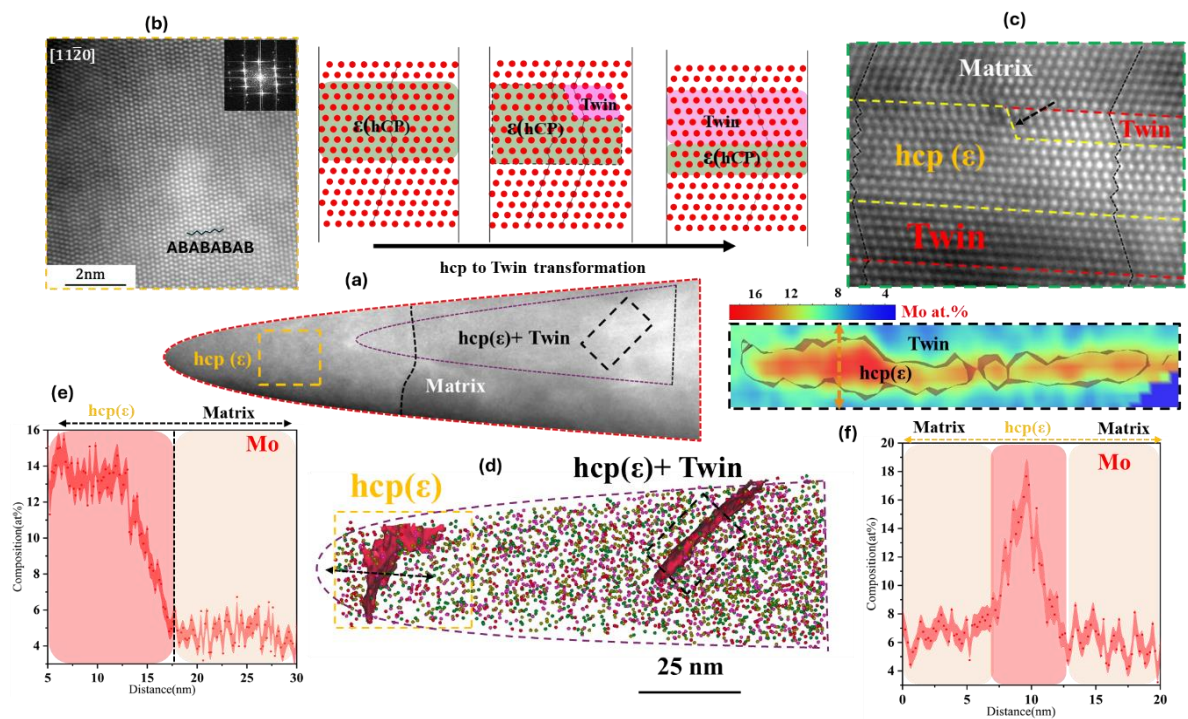

*Figure 4:* (a) An HAADF STEM image of a needle specimen taken from a deformed region of 65CRT alloy that has hcp ε nano-martensite at the top, separated by the interface from the fcc matrix at the bottom that also contains a laminated NT/NML structure. (b) An atomic-resolution HAADF STEM image taken from the top region of the needle along [11-20] zone axis of the hcp ε martensite showing ABABABA…. atomic sequence. (c) The atomic structure of the NT/NML laminated structure taken from the region indicated by the black-checked box shows the presence of a ledge between the hcp and fcc twin. (d) Atom probe reconstruction of the same needle specimen with the distribution of Co, Cr, Ni, and Mo atoms with two planar iso-compositional surfaces delineated by 12.5at.% of Mo (top) that separates ε martensite from the fcc matrix and 8 at.% Mo (bottom) corresponds to the ε martensite of the laminated NT/NML structure. (e) The Mo composition profile across the top interface shows Mo enrichment to the ε martensite up to 15 at.% (other profiles are shown in Supplementary Figure 8). (f) A 2-dimensional Mo compositional profile projected perpendicular to the viewing

*direction shows Mo-enrichment at the ε martensite of the laminated structure that goes up to ~16 at.%.*

Figure 4(d) shows the APT reconstruction of the same needle with the distribution of Co, Ni, Cr, and Mo atoms. The first ~50 nm from the top of the reconstruction is an NML structure with a Mo composition up to 16 at.% (Figure 4(e)). An iso-compositional surface of 10.5 at.% Mo clearly partitions as a phase boundary between the NML and the fcc matrix below. This solute partitioning to the NMLs structure is termed Solute-Partitioned NMLs (SP-NMLs). Other compositional profiles across the phase boundary reveal enrichment of Co up to ~43at%, while depletion of Ni and Cr up to ~28at% and 12at%, respectively, as shown in Supplementary Figure S5. The structural and compositional analysis of the NT/NML laminated structure found at the bottom of the tomogram is shown in Figure 4(c,f). As mentioned earlier, we observe a ledge, indicated by a black arrow, between the hcp structure of NML and the twinned region of NT that indicates a structural transition of ε martensite to twin. Figure 4(f) shows a 2D compositional map of Mo for the same laminated structure projected perpendicular to the viewing direction. The map reveals preferential partitioning of Mo to the ε martensite compared to the surrounding primary matrix and twinned matrix (NT). The profile taken across the NML shows the Mo composition is up to ~16 at.% in the ε hcp martensite region with respect to the matrix and the twinned region.

## 3.5 Thermal Stability of the Microstructure

Further, the 65CRT alloy with a larger amount of cold reduction shows higher microstructural stability at 600°C up to 100 hours (Figure 1 (c)) without hardening loss, as one could expect vis-a-versa. The hardness of the 45CRT alloy drastically reduces from 540 Hv to 470 Hv after tempering for 100 hours. The EBSD orientation maps indicate abnormal grain growth for the 45CRT (100 hours) alloy, while the grain size distribution in the 65CRT (100 hours) alloy remains stable, Figure 5(a,c). The grain growth in 45CRT alloy is driven by strain-induced boundary migration (SIBM) during the thermal treatment due to the gradient in strain energy induced by cold-rolling [39]. The Kernal Average Misorientation (KAM) map shows the grains exhibit a larger fraction of strain-softened areas in 45CRT, whereas a very low fraction of these areas in 65CRT alloy. Since the grain boundaries (GBs) are the primary source for the nucleation of twinning partials, their migration also results in the rapid thickening of NTs to micro-twins [40], as shown in Figure 5(b) with several thick micro-twins present in the strain-softened areas of 45CRT alloy after 100 hours of tempering [41,42]. However, we couldn't see similar thick micro-twins in 65CRT alloy (Figure 5(d)). This indicates the SP-NMLs in the microstructure inhibit the grain boundary migration by pinning the GBs in 65CRT alloy [29]. Hence, the microstructure still retains the hardness with the SFs, NTs and NMLs inside the grains.

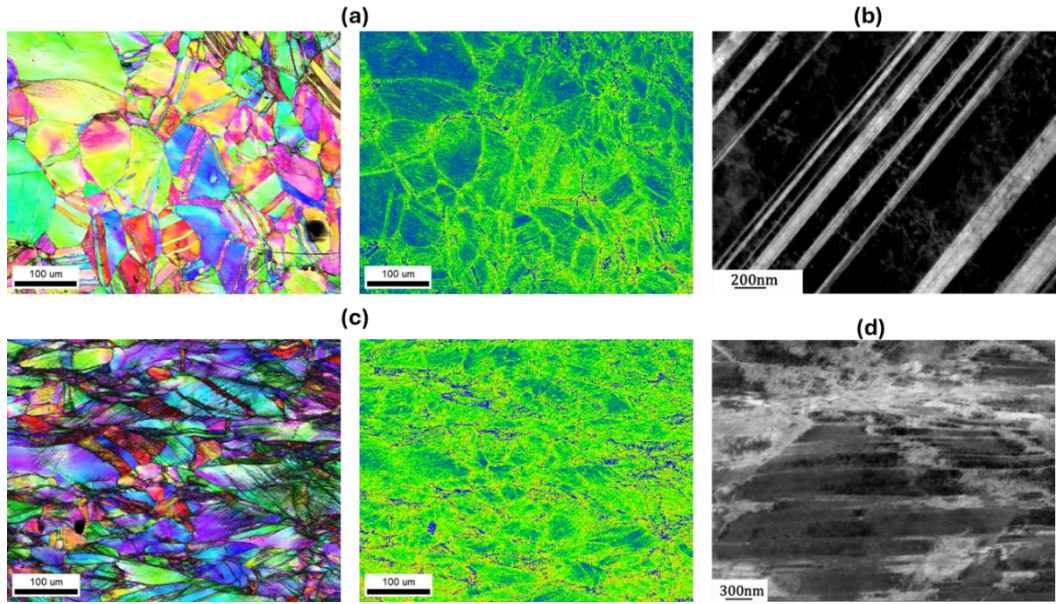

*Figure 5:* (a) EBSD orientation and KAM maps of 45CR after 100 hours of tempering showing large fraction of strain-softened areas. (b) an electron-channeling-contrast image (ECCI) ECCI image from a strain-softened area showing thick micro-twins. (c) Similar, EBSD and KAM maps shows the uniform high strain distribution across the microstructure. d ECC image showing deformation structure is retained.

## 4. Discussion

The experimental results show that the amount of cold working governs the type of deformation structures to be formed in Ni-Co-Cr-Mo alloy. Subsequent tempering at 600°C results in further hardening of the alloy that is due to the solute segregation/partitioning to the structures that are formed during cold rolling. Interestingly, the alloy with a larger degree of cold working shows higher microstructural stability during tempering, as otherwise one could expect vis-a-versa. This is attributed to the presence of solute-partitioned nano-martensite-laths (SP-NMLs) in addition to the solute-segregated SFs and nano-twins (NTs) in the microstructure of 65CRT alloy. In the subsequent section, we discuss the experimental results centered on the obtained mechanical properties.

### 4.1 Formation of deformation-induced structures after cold rolling

In face-centered-cubic (fcc) based alloys, the strain can be accommodated via either or a combination of dislocation slip, stacking faults (SFs), twinning, and hexagonal-close-packed (HCP) martensitic transformation (FCC→ ε-HCP) during the plastic deformation. The activation of these mechanisms is found to be directly related to the intrinsic stacking fault energy $\gamma_{iSFE}$ of the alloy [43], i.e., the plastic deformation is dominated by twinning ($\leq$ 20-40 mJ/m²) and hcp martensitic transformation (($\leq$ 20 mJ/m²). Otherwise, the deformation predominantly occurs by glide of 1/2<110> full dislocations [43]. The formation of an SF is facilitated by the movement of 1/6<112> partial dislocations on the (111) plane, which act as an embryo for other planar defects transformation. The atomic structure of these SFs is equivalent to a two-layer hcp structure (called intrinsic stacking fault (ISF)). These SFs can transform into an ESF or twin by the consecutive passage of the same partial on {111} planes adjacent to the SF that has an activation barrier (AB) for its partial nucleation ($E_{tw}^{fcc}$) [44]. Similarly, these SFs will transform into martensite if the partial dislocation passes on an alternate {111} plane, which has an AB for the partial nucleation ($E_{mar}^{fcc}$) on one layer away from the SF. Supplementary Figure S6 shows the schematic of these transformations.

The SFE of the current alloy is measured experimentally via weak-beam TEM (see Supplementary information S7), having a low value of ~ 12mJ/m$^2$. This indicates that the alloy can accommodate plastic strain via forming SFs or martensite transformation (NML). The microstructure analysis of 45CR alloy shows SFs and NTs, while 65CR shows SFs with a higher fraction of NTs and NMLs. This shows a higher AB for martensite nucleation (NML) than NTs, as reported in [45–48]. The ABs of NTs and NMLs can be calculated by the generalized stacking fault energy curve [44,49] and are also a function of lattice frictional stress associated with the lattice distortion of the concentrated solid solution matrix [45,50–52]. These lattice frictional stress (or ABs) can be overcome locally by the stress concentration produced during the deformation, i.e., the larger the amount of deformation/distortion introduced into the matrix, the higher the fraction of NTs and NMLs in the microstructure [51–54]. Figure 7(a) shows a higher fraction of NTs and NMLs in 65 CR alloy than the 45CR alloy (note that no NML was present in 45 CR alloy). Supplementary Figure S8 also shows the microstructure of 25CR alloy where only SFs were observed throughout the microstructure, which underpins the importance of the degree of cold reduction/deformation in overcoming the AB for the formation of different defect structures. A further strain can also be accommodated by the deformation inside the NMLs by the movement of partial dislocation a/3<10$\bar{1}$1> that will change the hcp sequence of ABABAB…. To Twin (FCC) ABCABC…. Sequence [48]. This results in the laminated NT/NML structures in the alloy. Figure 4(c) shows a high-resolution image centered on a laminated structure with a ledge across the hcp and twin structure, indicating the transformation. The schematic representation is also shown, known as the transformation-mediated twinning (TMT) process [48].

## 4.2 Solute segregation/partitioning to the deformation-induced structures after tempering

Figure (3-4) reveals evidence of solute segregation/partitioning (Mo/Co) to the deformation-induced defect structures on tempering (600°C for 4 hours) of the alloy. This is due to the relatively higher stability of Mo/Co in the hcp structure [55]. Hence, on thermal activation (tempering), Co and Mo prefer to occupy the hcp lattice structure compared to the surrounding fcc lattice, which is driven by the reduction of overall system Gibbs free energy. Reducing twin boundary energy also drives Mo/Co segregation to NTs [37]. The compositional profiles in Figure 3(d) show Mo/Co enrichment peaks at the two NTs of nearly ~ 2-4 nm thin. The profiles cannot resolve into separate peaks at the twin boundaries for each NT due to the larger compositional width relative to the thinness of the NTs [56]. The nature of enrichment (Mo/Co) to the SFs and NTs is similar for both 45CRT and 65CRT alloys. Compositional analysis of ε-hcp NML in 65CRT alloy shows an enrichment of Mo up to ~16 at.%, indicating a higher ability of NML to accommodate Mo as compared to the SFs and NTs.

Since the local structure of SFs and NMLs is hcp, which is generated during plastic deformation and coherently embedded in the fcc matrix, the local stress state and thermodynamic boundary conditions are identical. Hence, this cannot lead to the difference in the composition of the two phases, which is different from our experimental observations. It is to be noted that the hcp atomic layers of SFs are 1-3 thick as compared to relatively thicker NMLs (15 layers to a few nm) in the surrounding cubic environment of fcc atomic structure. The finite thickness of SFs with hcp atomic structure leads to the interactions between the second nearest neighbours located in the SF and adjacent fcc matrix [32,57]. More specifically, the atoms in the SF have both hcp and fcc environment through second nearest-neighbour interactions. However, the atoms inside NMLs have an hcp environment through second nearest-neighbour interactions. Hence, the composition of Mo at SF compromises the higher value for NML and the lower value in bulk fcc. Additionally, the Monte Carlo simulations show that the reduction in SF energy occurs up to a critical composition of Mo beyond which the local lattice gets distorted, which increases the intrinsic strain energy, inhibiting further accumulation of Mo atoms [32] at SF. Simulations also reveal that the segregated Mo atoms prefer to attract Co as compared to Ni and Cr at the SFs. Additionally, the binary phase diagram between Co and Mo shows a line compound with a stoichiometry of $Co_3Mo$ equilibrium bulk phase having the hcp-based $DO_{19}$ ordered structure. Thermodynamic phase diagram calculations [58] by Thermocalc for $(CoNi)_{100-x}$ - $Mo_x$ also show a two-phase structure of fcc + hcp at 600°C for Mo composition ~> 8 at.%. Since, in the NMLs, Co and Mo both are enriched, one could expect that tempering for a longer time, the structure of NML can transit from hcp to $DO_{19}$ ordering. However, we don't see any structural transition on tempering even up to 100 hours, which might be due to the presence of Ni (~ 30at.%) that is known to increase the enthalpy of formation ($\varDelta H_f$) of $D0_{19}$ ordered structure [59]. In the case of $L1_2$ ordered phase in Co-Ni base superalloys, similar segregation behavior of Mo/Co (much lower than 25 at.%) was observed to the creep-induced SFs, which has locally $D0_{19}$ ordered structure coherent with the surrounding $L1_2$ structure [60]. The behavior was associated to the local phase transformation strengthening that is shown to influence the overall creep strength of the alloy. This is different from the present case since the segregation of Mo/Co solutes occurs at the disordered hcp structure of SFs/NMLs (twin boundaries of NTs).

Based on the above structural and compositional measurements of defect structures in the alloy after cold rolling and tempering, respectively, we observe the following sequence of microstructural evolution: Cold-rolling induces (1) structural transition of local fcc to hcp structures in the form of SFs, NTs in 45CR alloy while additional formation of NMLs in 65 CR alloy. (2) nano-laminated structures (NT/NML) in 65CR alloy by the transition of hcp structure of NML to NT structure via TMT mechanism. Tempering induces (3) segregation/partitioning of Mo/Co to the defect structures whose composition is limited by their local structural dimension and distortion at the hcp structure in the fcc matrix.

### 4.3 Influence of deformation-induced structures on the tensile properties

The 45CR/65CR samples exhibit a high 0.2% YS (up to ~1.5 GPa) at room temperature, which enhances up to ~2 GPa after tempering at 600°C attributed to solute segregation/partitioning by tempering to the defect structures produced during cold rolling. The onset of plastic deformation during loading involves generations of mobile dislocations and the movement of the pre-existing dislocations. Interfaces in the microstructure, such as GBs, phase boundaries, twins, and planar defects, act as a source for the nucleation of new dislocations and also provide resistance to dislocation motion during plastic deformation. Solute segregation to these interfaces creates an additional stress field and increases the difficulty in dislocation nucleation as well as the stress required for the dislocation glide. Here, we will continue discussing the defect structures of solute-segregated SFs, NTs, and SP-NMLs. Figure 6 shows the atomic structure and corresponding lattice strain maps of SFs in 65CR and 65CRT alloys generated by the procedure described in [61]. The degree of localized lattice strain at the ISF is higher after solute segregation, i.e., after tempering. This is attributed to segregating larger atomic size Mo atoms at the fault plane [32]. As mentioned earlier, Monte Carlo simulations also show a strong tendency of increased lattice distortion by Mo atoms on the fault planes. Hence, during loading, the strain field of the moving dislocations interacts with the higher localized strain of the solute-segregated defect structure and hence experiences an extra resistance for their glide.

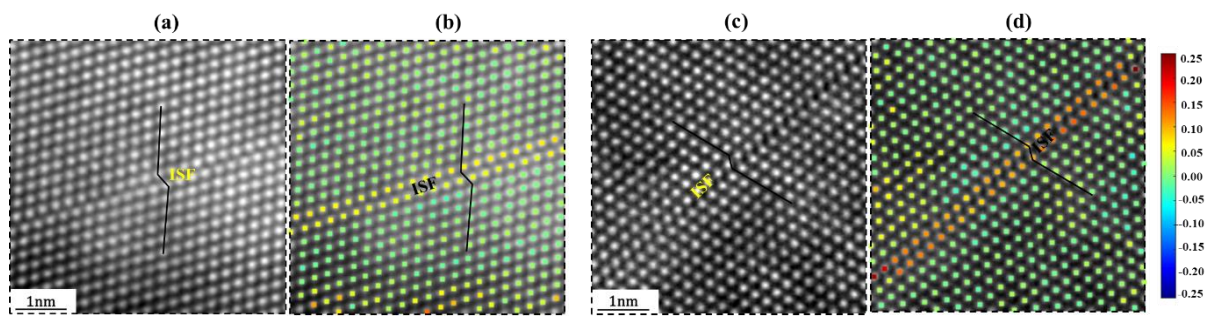

*Figure 6:* *(a, c) HR-STEM image of 65CR and 65CRT samples showing the atomic structure of an ISF, respectively. (b, d) lattice strain mapping across the SF in 65CR and 65CRT, respectively. The localized lattice strain surrounding the SF is slightly higher in the 65CRT sample due to the segregation of Mo atoms.*

In addition, the nature of solute segregation to the defect structures, i.e., the enrichment of Co/Mo indicates a larger localized number of Co-Mo bonds, which has larger negative values of enthalpy of mixing (~ 1.0 eV/solute atom) [62] in the local hcp lattice. This can resemble a short-range ordering of Co-Mo bonds, and hence, during tensile loading, higher stresses are required for the dislocations to cut through these bonds and facilitate a higher YS [63,64]. Atomic-scale compositional analysis revealed different amounts of Mo segregation between the defect structures with Mo composition 7-8 at.% at SF and NTs and up to 15-17 at.% in NMLs. For 45CRT alloy, the increase in 0.2% YS results from resistance to dislocation motion by the increased localized strain and the Co-Mo bonds at the fault planes or twin boundaries. In 65CRT alloy, the additional strengthening contribution comes from the higher number of Co-Mo bonds in NML, with the local hcp atomic structure having a relatively limited number of slip systems compared to the surrounding fcc atomic structure. This requires higher stresses for the movement of dislocations across from the fcc matrix and slip transfer across these fcc/NMLs phase boundaries for the dislocation glide, respectively. As a result, the 65CRT alloy exhibits a larger increase in 0.2% YS (~500 MPa) after tempering compared to the 45CRT alloy (~200 MPa).

## 4.4 Effect of Cryo-rolling on the tensile properties and microstructure stability

Based on the above discussion, we conclude that if a higher fraction of NMLs is introduced in the microstructure, superior strength and stability can be achieved at ambient and high temperatures. The propensity of the fraction of these defect structures can increased by reducing the activation barrier (AB) for their nucleation in the matrix. A commercially viable process is cryo-deformation, where the AB drastically reduces with decreasing temperatures [44,48,65,66]. It also reduces the critical shear stress required for twinning and martensite formation and can promote plastic strain accommodation by extensive twinning and martensite formation. This is shown successfully in several multi-component solid solution alloys with low SFE values during tensile testing at cryo temperatures exhibiting higher tensile strength and toughness [67].

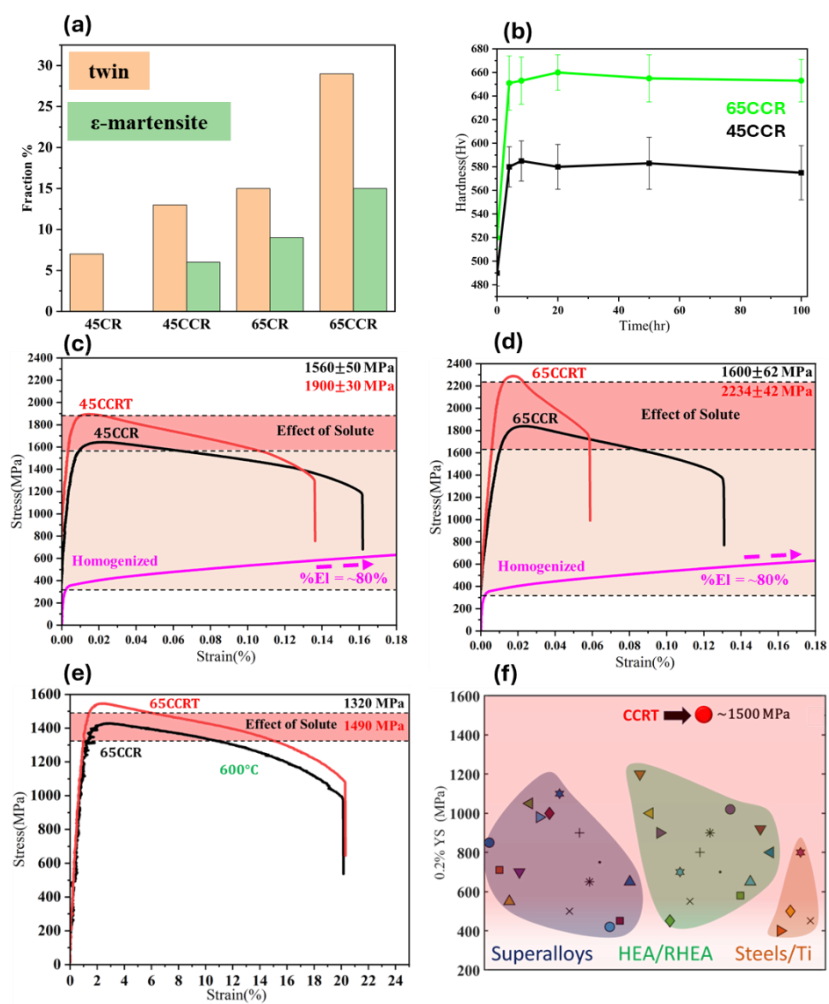

*Figure 7: (a) Comparison of Fraction of NTs and NMLs between 45CR, 45CCR, 65CR and 65CCR alloys. (b) Hardness vs tempering time for the cryo-rolled alloys (45CCR and 65CCR). (c) Tensile engineering stress-strain plots after cryo-rolling and tempering (4 hours). (d) A plot showing the effect of solute segregation/partitioning to the defect structures produced during cold- and cryo-rolling (45% and 65% thickness reductions). (e) High temperature (600°C) tensile engineering stress-strain plot for the 65CCRT. (f) plot showing the comparison of 0.2% YS at 600°C of 65CCRT alloy with other precipitation hardenable superalloys, high strength steels, and multi-principle element alloys (fcc-based and refractory-based).*

From this knowledge, cryo-rolling with 45% and 65% reductions (45CCR and 65CCR) was carried followed by tempering at 600°C (45CCRT and 65CCRT). Figure 7(a) shows the comparison of the fraction of defect structures between the cold- and cryo-rolled alloys. Most importantly, the 45CCRT alloy also contains ~6% SP-NMLs, while the fraction increased to ~15% (from 7%) in 65CCRT alloy. Similarly, the fraction of NTs also increased by cryo-rolling of both alloys. Figure 7(b) shows the hardness variation vs tempering time up to 100 hours. Cryo-rolled alloys show higher hardness values after 4 hours of tempering than cold-rolled alloys. Interestingly, the hardness of 45CCRT alloy doesn't decrease until 100 hours, unlike 45CRT alloy, which showed a drastic reduction in the value. Hence, this indicates that the stability is due to the presence of ~ 6% SP-NMLs in the microstructure for 45CCRT alloy, which was not formed during cold rolling. Also, in the case of 65CCRT alloy with a higher fraction of SP-NMLs, the microstructure is stable up to 100 hours of tempering. The tensile curves, shown in Figure 7(c-d), reveal a 0.2% YS of ~1.9 GPa for 45CCRT alloy while a 0.2% YS of ~2.2 GPa for 65CCRT without any change to the elongation to fracture as compared to cold-rolled alloys. This shows that cryo-rolling can be used as a processing step to attain a higher 0.2% YS for a given percentage of thickness reduction without reducing the elongation to fracture values. Tensile testing at 600°C, the 65CCRT alloys show an extraordinary 0.2% YS of ~1.5 GPa with a ~19% elongation to fracture, as shown in Figure 7(e). This high value of YS, at 600°C, is superior to all the existing precipitation hardenable high-temperature superalloys, high-strength steels, and multi-component fcc-based and refractory alloys. Figure 7(f) shows an Ashby plot between the YS and %El to fracture at 600°C for comparison that reveals a unique place in the plot (composition of alloys is shown in Supplementary Figure S9).

### 4.5 Comments and discussion on the previous works on the alloy

Based on the above discussions, the present correlative microscopy work reveals several unknown facts that contribute to the extensive hardening of the alloy and provide pathways that hold on alloying and processing to design ultra-strong structural alloys.

To counter recent works (as mentioned in the introduction) using atom probe tomography, Figure 8(a) shows a high-resolution EBSD orientation map of the 65CR alloy, revealing several elongated nano-grains with misorientations not larger than 8 to 10 degrees (Figure 8(b)). A site-specific needle specimen (red dashed enclosure) was extracted by the in-plane method [68] for atomic scale compositional analysis across these grains. Figure 8(c) shows the APT reconstruction with the distribution of Co (dark yellow), Ni (green), Cr (pink), and Mo (red) atoms. An elongated feature extending along the tip direction was reconstructed using a Boron (B) 0.2at% iso-compositional surface that corresponds to the grain boundary highlighted in Figure 8(a) by a white arrow. Additionally, several other planar features are also reconstructed using Cr 27 at.% iso-composition (pink color) surface. These planar features extend across the tip and terminate at the GB, and the regions enclosed between these features are the elongated nano-grains. Compositional profiles across the GB (Figure 8(d-e)) and the nano-grains (Figure 8(f-g)) show clear evidence of Cr, a slight amount of Mo, and B while, surprisingly, depletion of Co along with the Ni. The segregation behavior of solutes exactly matches the recently published data [33–35], which claims these are stacking faults (SFs). Additionally, no evidence was provided for their claims that the shown planar features in their APT data are SFs. Note that the data in Figure 8 is from the cold-rolled alloy (without tempering), indicating the occurrence of deformation-induced grain-boundary (GB) segregation due to the dynamic interaction of solutes with the gliding dislocations [69] during deformation.

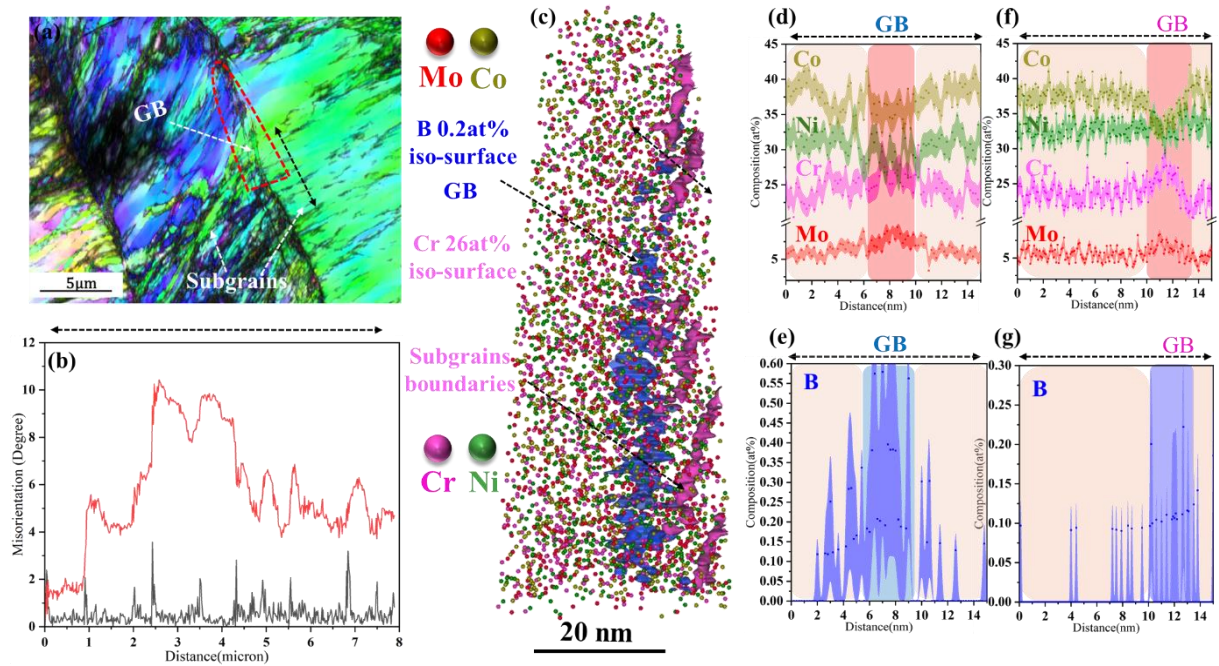

*Figure 8:* (a) high-resolution EBSD map of the 65CR sample showing some elongated grains and nano sub-grains. (b) mis-orientation profile along the black dashed line indicated in a. (c) APT reconstruction of a needle specimen, extracted from the region marked and enclosed by the red dashed box, highlighting the elongated GB by B iso-surface (blue) and sub-grain boundaries by Cr iso-surface (magenta). (d-e) composition profile across the elongated GB. (f-g) composition profile across the sub-grain boundary.

Similarly, after tempering, a correlative analysis of a GB in 65CRT alloy is shown in Figure 9(a). A TEM-DF image of a needle specimen was taken using a diffraction spot that corresponds to the grain, which is highlighted in Figure 9(b). The same specimen was field evaporated, and the reconstruction is shown in Figure 9(c) with a 28 at.% iso-composition surface of Cr that represents the same GB (the reconstruction is shown in the edge-on condition of the GB). The compositional profiles (Figure 9(d-e)) show the same segregation behavior as for the GBs in cold-rolled conditions, i.e., enrichment of Cr, Mo, and B while depletion of Co and Ni. These results indicate that the increment of hardness after tempering cannot be attributed to the GB segregation. To validate further if the SFs in the cold-rolled alloys have segregating solutes on them, a needle specimen was prepared from the 65CR alloy for the compositional analysis. Figure 9(f) shows a STEM image of the specimen containing SFs in an edge-on condition passing across the needle (indicated by yellow arrows). The same needle was field evaporated, and Figure 9(g) shows the reconstruction. The composition profiles and 2D compositional map of Mo and Cr, Figure 9(h-i), reveal no tendency of solutes segregating to the faults, i.e., the composition is uniform across the specimen.

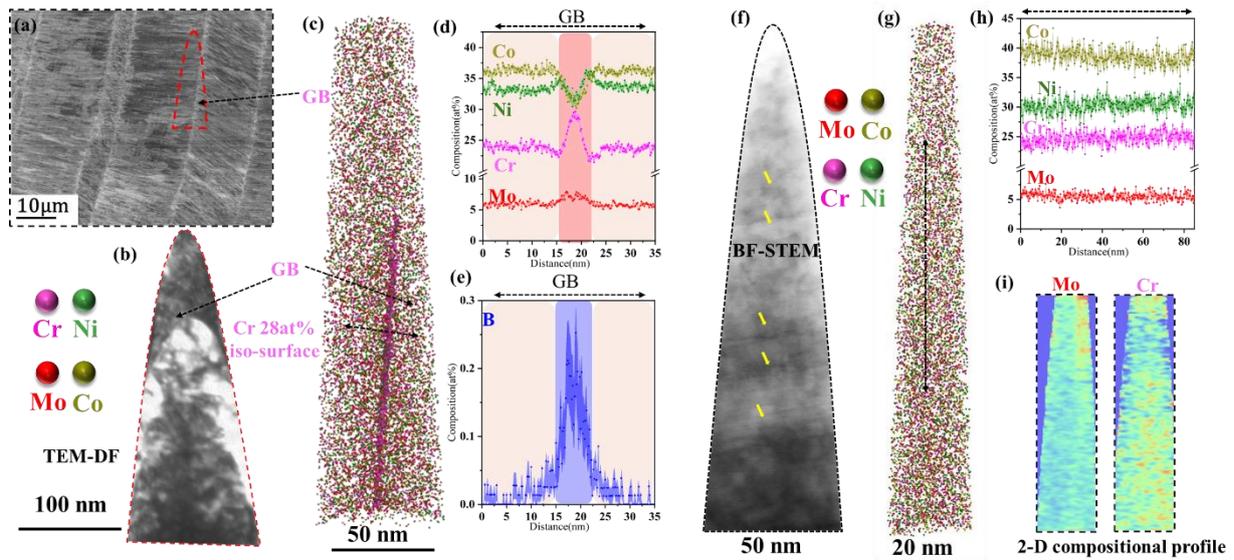

*Figure 9:* *(a) BSE-SEM micrograph taken from a 65CRT sample showing elongated GBs. (b) DF-TEM image of a needle specimen extracted from the region marked and enclosed by a red box area confirms the presence of a GB. (c) APT reconstruction displaying the distribution of Co (dark yellow), Cr (magenta), Ni (green), and Mo (red atoms) with a planar iso-composition surface delineated by 27at.% of Cr t, corresponding to the GB. Composition profile across the GB of (d) Co, Cr, Ni, and Mo (e) B. (f) HAADF-STEM image of a needle specimen from the 65CR sample showing several SFs in the edge-on condition. (g) APT reconstruction of the same needle specimen with the distribution of Co, Cr, Ni, and Mo atoms. (h) Compositional profiles of Co, Cr, Ni, and Mo along the dashed line reveal uniform composition. (i) Corresponding 2-D compositional maps Cr and Mo showing uniform distribution.*

In the present work, we show the formation of SFs and nano-twins in the 45CR sample, while in the 65CR alloy, additional nano-martensite-laths (NMLs) are observed. Hence, the hardening of the alloys after tempering is associated with the segregation/partitioning of solutes (Mo/Co) to the defect structures formed during rolling. This indicates that in addition to the Suzuki strengthening, the presence of SP-NMLs in the microstructure results in the extensive strengthening of the alloy, which has not been found or discussed in previous works.

## 5. Conclusions

The experimental results reveal a phenomenon of atomic-scale solute segregation/partitioning to the structures produced during cold-working of a low SFE alloy increases the YS beyond 2GPa. These are termed solute-segregated structures that are SFs (stacking faults), NTs (nano-twins), and SP-NMLs (Solute-Partitioned Nano-Martensite-Laths), Figure 10(a). Among these, SP-NMLs are also shown to increase the high temperature stability of the microstructure (up to 600°C). The type and fraction of defect structures are tuneable based on the degree and temperature of cold deformation. This is demonstrated in a Ni-Co-Cr alloy with 5at.% Mo as a solute where a cryo-rolling by 65% cold-reduction followed by tempering at 600°C results in a tensile YS of ~ 2.2GPa with an elongation to fracture ~6%. At 600°C, the alloy shows an exceptional YS value of ~1.5 GPa with the elongation to fracture up to ~ 18%, which is much higher than conventional precipitation hardenable alloys, steels, and other high entropy alloys.

Figure 10(b) shows the comparative plot of YS vs processing conditions and the effect of solute segregation/partitioning to the deformation-induced structures in the alloy. Based on the exploration and revelation of these unique defect structures in fcc-based Ni-Co-Cr-Mo alloy, we propose that the strategy can be effective in other low SFE alloys with suitable solutes (that either reduces the SFE and/or lower the AB of the defect structures), can open avenues for designing ultra-strong structural materials.

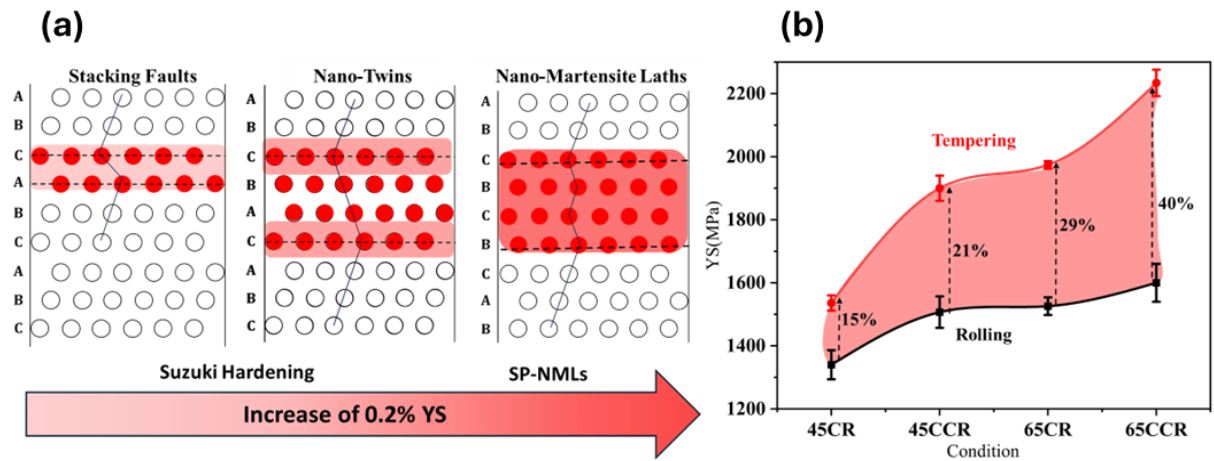

*Figure 10:* (a) Schematic representation of the solute segregation/partitioning to the deformation-induced structures in the alloy. (b) A plot showing the effect of the degree of cold-working followed by tempering on the YS of the alloy.

**Acknowledgments**

We gratefully acknowledge the Advanced Facility for Microscopy and Microanalysis (AFMM), Indian Institute of Science, Bangalore, for providing access to FIB, TEM, and APT facilities. SKM and AG acknowledge the financial support from ISRO-IISc STC cell and MPG-IISc Partner Group. AG is also thankful for the PMRF fellowship.

# Supplementary Information

**Exploiting solute segregation and partitioning to the deformation-induced planar defects and nano-martensite in designing ultra-strong Co-Ni base alloys**

Akshat Godha[1*], Mayank Pratap Singh[1], Karthick Sundar[1] Shashwat Kumar Mishra[1], Praveen Kumar[1], Govind B[2], Surendra Kumar Makineni[1*]

[1]Department of Materials Engineering, Indian Institute of Science Bangalore – 560012

[2]Materials and Metallurgy Group, Vikram Sarabhai Space Centre, Thiruvananthapuram – 695022

***Corresponding Authors:** akshatgodha@iisc.ac.in; skmakineni@iisc.ac.in

**S1: Microstructure after deformation (cold rolled): 45CR and 65CR alloys**

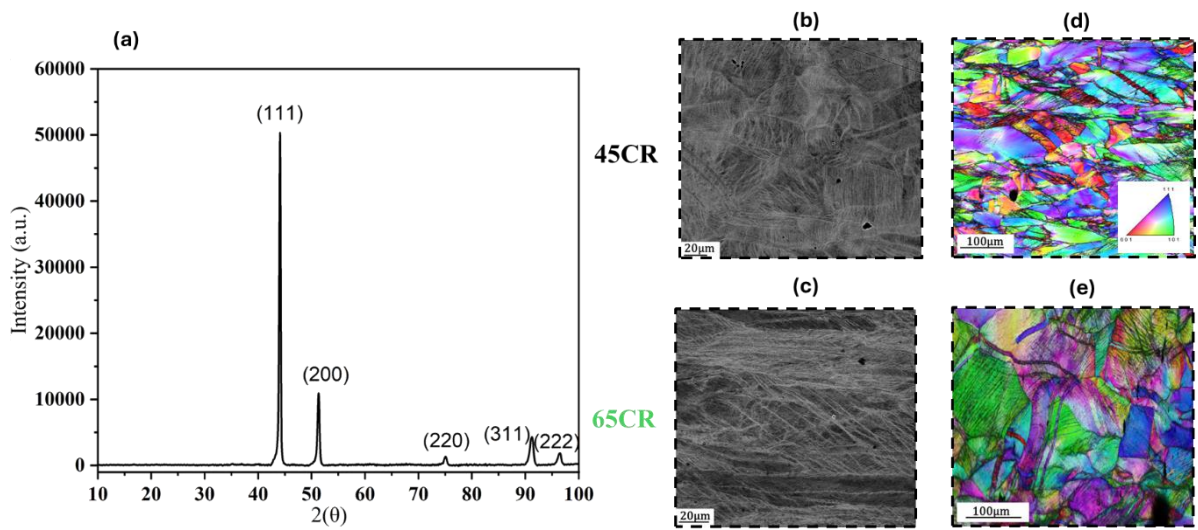

**Supplementary Fig. S1: a** X-ray Diffraction pattern for the homogenised Ni-33Co-27Cr-5Mo alloy showing all the peak intensities corresponding to face-centered-cubic (fcc) structure. **b** Low magnification back-scattered-electron (BSE) images for the 45CR and **c** 65CR alloys showing bright and thick shear bands. **d** Electron-back-scattered-diffraction (EBSD) grain orientation maps for 45CR and **e** 65CR alloys. High magnification electron chanelling contrast (ECC) images from **f** 45CR and **g** 65CR alloys.

**S2: Defect structures in 45CRT and 65CRT alloys (after tempering at 600°C)**

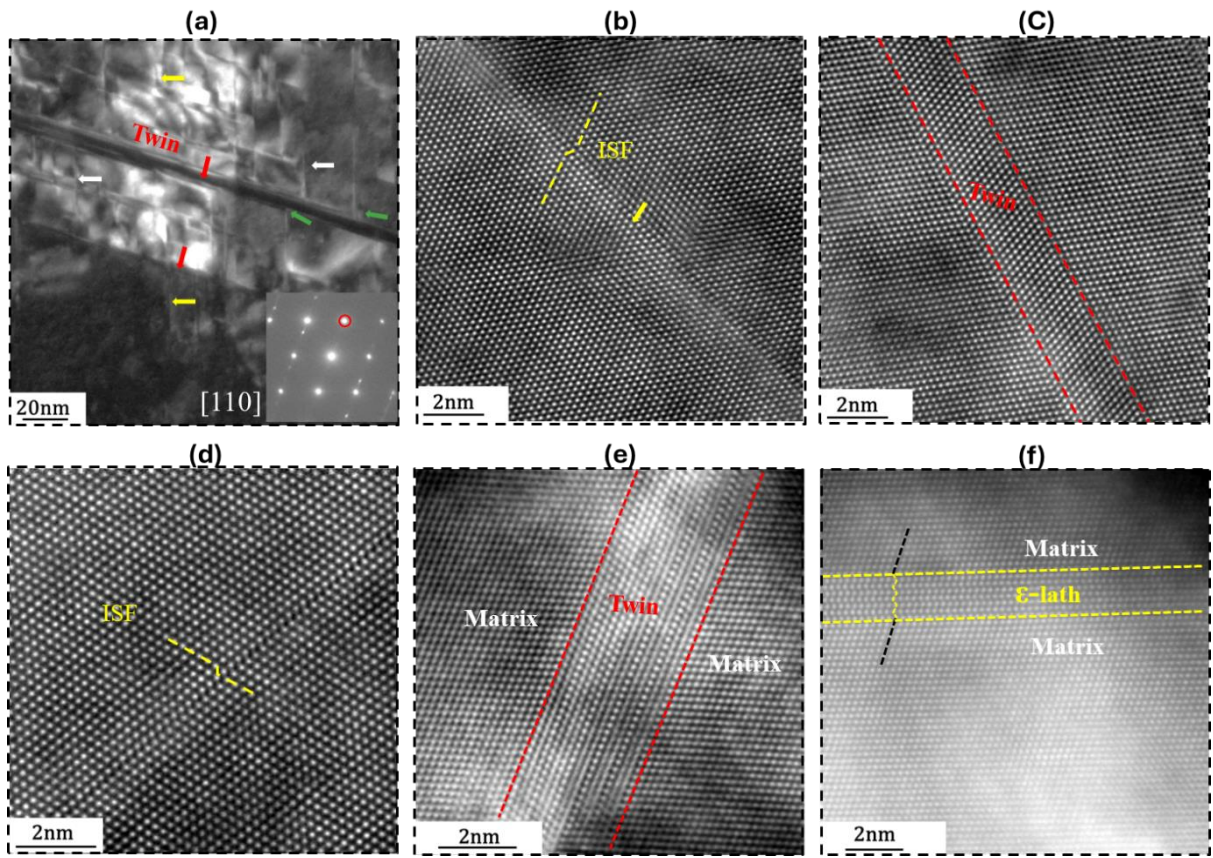

**Supplementary Fig. S2**: a. BF-TEM images of the 45CRT sample captured along the [110] zone axis, highlighting SFs (yellow arrow), NT (red arrow) and L-C locks (white arrow).) HAADF-STEM images centered on b. SF and c. NT. Similarly, HAADF-STEM images of 65CRT sample showing d. ISF/ESF. e. NT. f. NML

## S3. 1D and 2D compositional profiles across the SF from the APT reconstruction shown in Fig. 3b

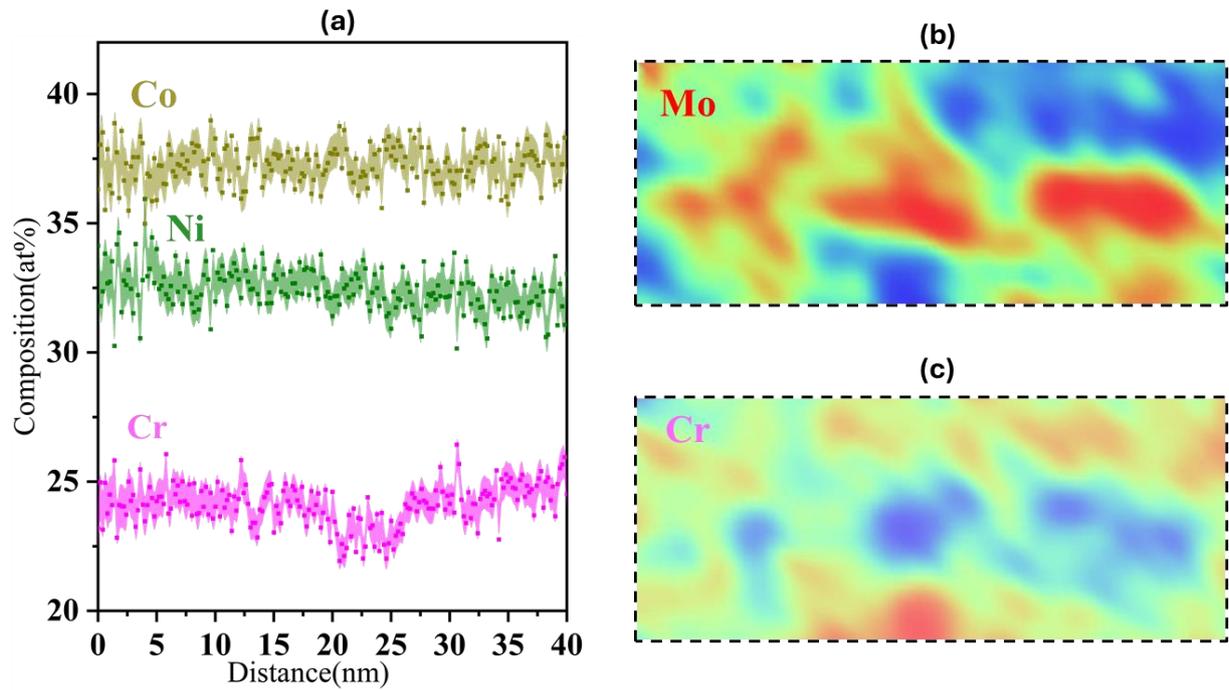

***Supplementary Figure S3****:* 1D compositional profile of **a** Co, Ni and Cr, **b** 2D compositional maps of Mo and **c** Cr along the SF plane.

## S4. Atomic scale compositional analysis of SFs and NTs in 45CRT alloy

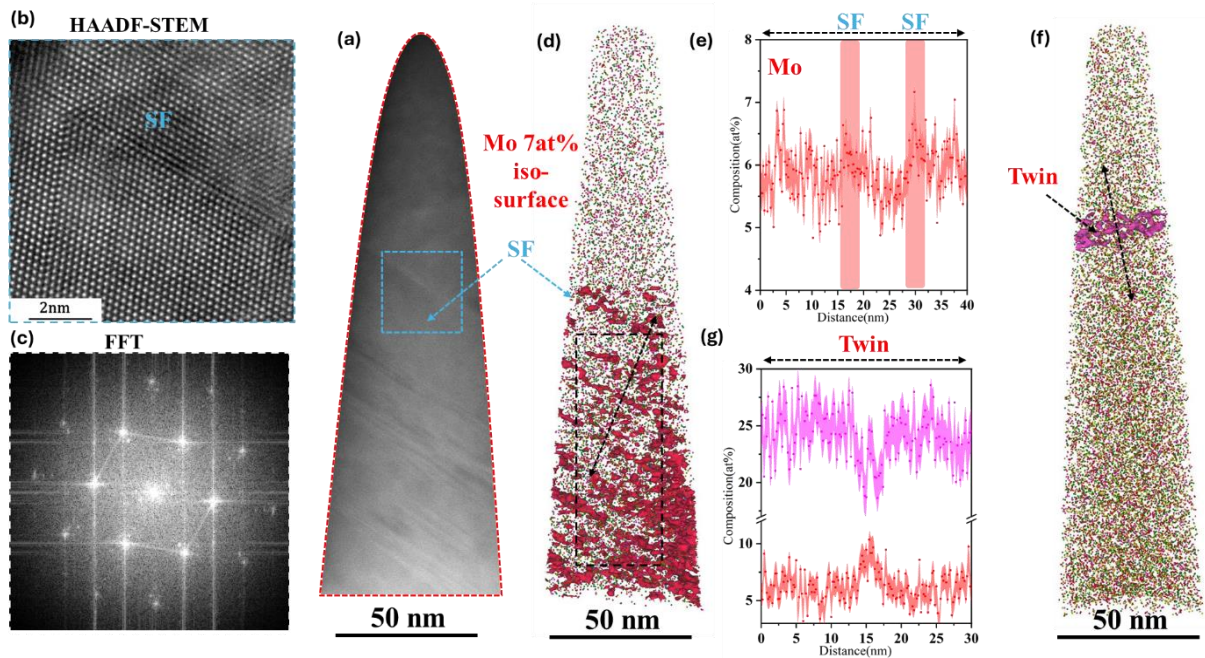

**Supplementary Fig. S4**: **a** HAADF-STEM image of a needle specimen from the 45CRT sample showing several SFs. **b** HR-STEM image showing the atomic structure of an SF. **c** shows the corresponding FFT, **d** APT reconstruction of the same needle specimen with the distribution o Co, Cr, Ni and Mo atoms Ni, showing a planar iso-composition surface delineated by 7at.% of Mo, corresponding to the SF. **e** Compositional profiles of Mo across the SF. **f** APT reconstruction of the twin containing needle specimen with the distribution of Co, Cr, Ni and Mo atoms Ni showing a planar iso-composition surface delineated by 19at.% of Cr, corresponding to the twin. **g** Compositional profiles of Cr and Mo across the SF.

## S5. Compositional profiles across the NML structures shown in the Fig. 4

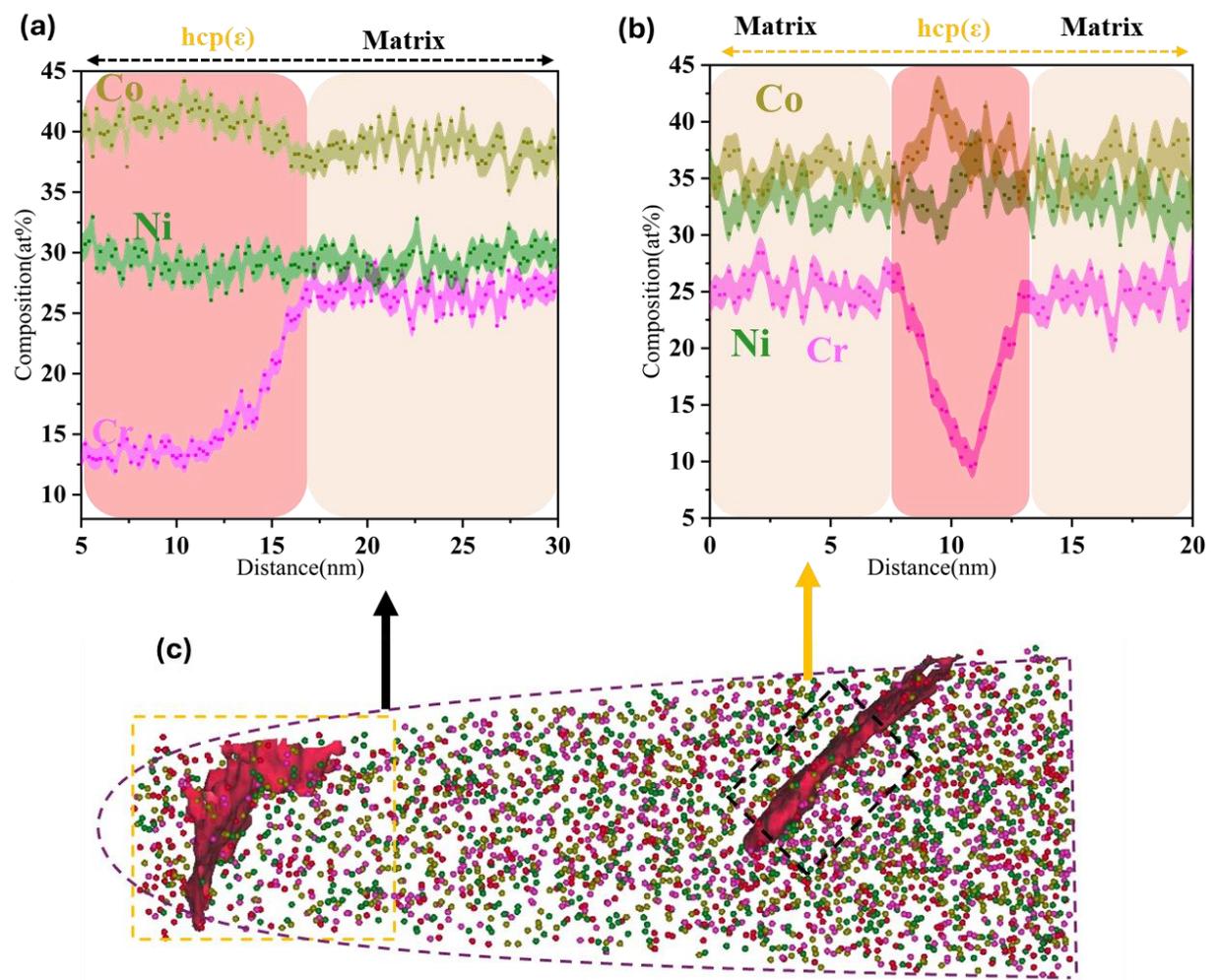

***Supplementary Figure S5***: 1D compositional profile of Co, Ni, Cr across **a** top hcp(ε), **b** bottom hcp(ε). **c** Atom probe reconstruction shown in Fig. 4 d.

## S6. Structural transitions from FCC to SFs, SFs to hcp ε, SFs to twin and hcp ε to twin

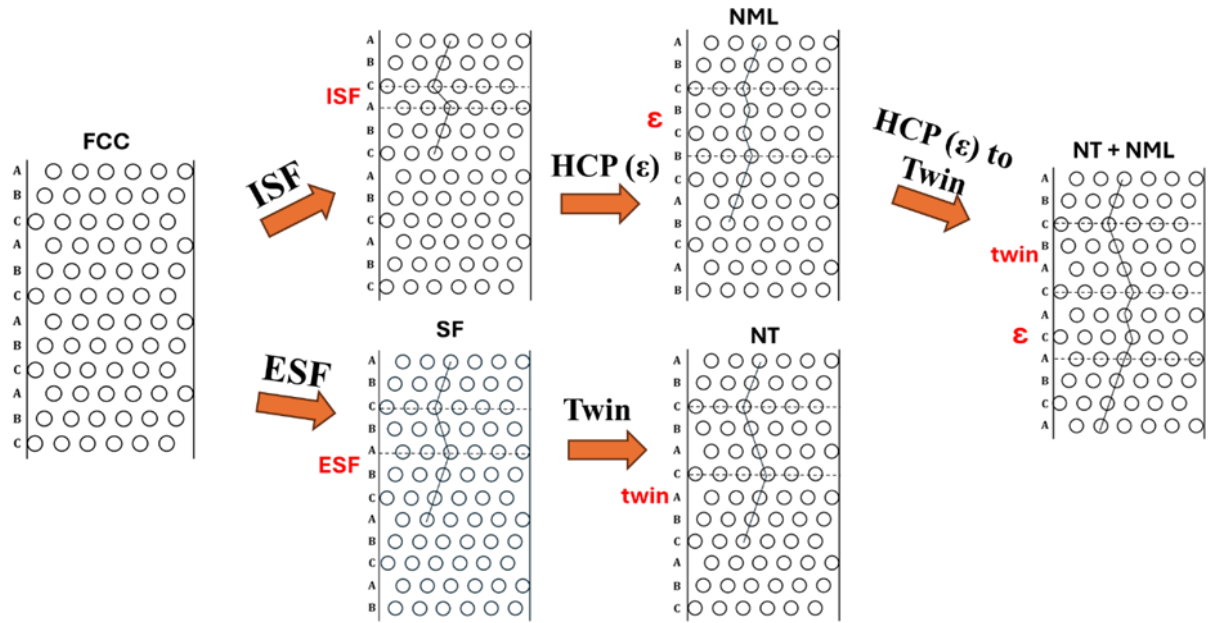

**Supplementary Fig. S6:** Schematic representation of the deformation induced structural transformation from FCC to SFs, SFs to hcp ε, SFs to twin and hcp ε to twin

## S7. Estimation of Stacking fault energy by weak-beam TEM

The stacking fault energy (SFE) was calculated by measuring the separation of two partial dislocations based on the following equation [1]:

$$\gamma = \frac{\mu b_p^2}{8\pi d}\left(\frac{2-\vartheta}{1-\vartheta}\right)\left(1 - \frac{2\vartheta \cos 2\alpha}{2-\vartheta}\right)$$

Where $\mu$ is the shear modulus, b is the magnitude of the Burger vector of partial dislocations and $\vartheta$ the Poisson's ratio. Values of the shear modulus of 81GPa and Poission's ratio of 0.325 were used from the study [2]. $\alpha$ is the angle between the Burger vector of the perfect dislocation and dislocation line. The magnitude of the burger vector is measured to be 0.146nm. Figure 4a shows a g-3g weak beam dark-field (WBDF) micrograph of a dissociated dislocation with g=220. The average separation distance between these two partials is measured to be ~9nm. Using the above equation, the average SFE is estimated to be 12.5±1.5 mJ/m$^2$.

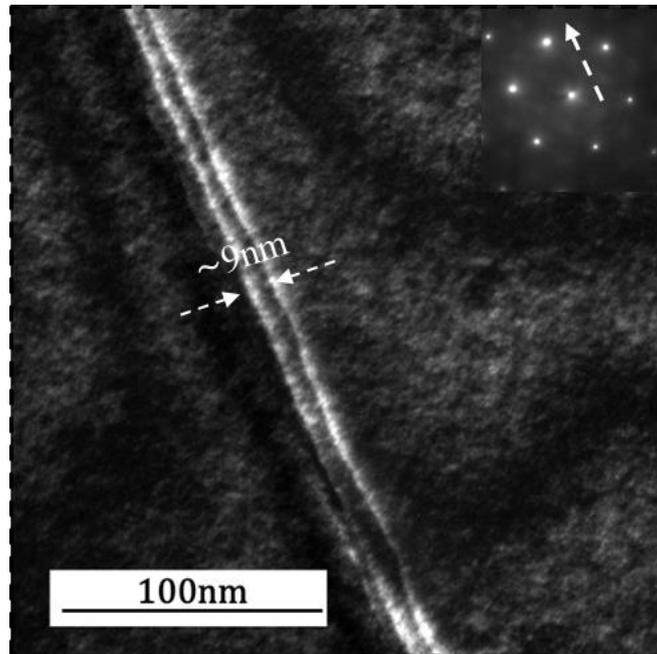

**Supplementary Fig. S7** : *Weak beam dark field (WBDF) image of a dissociated dislocation, highlighting the separation between two partial dislocations.*

**S8. Bright-field TEM images of 25CR alloy showing stacking faults (SFs)**

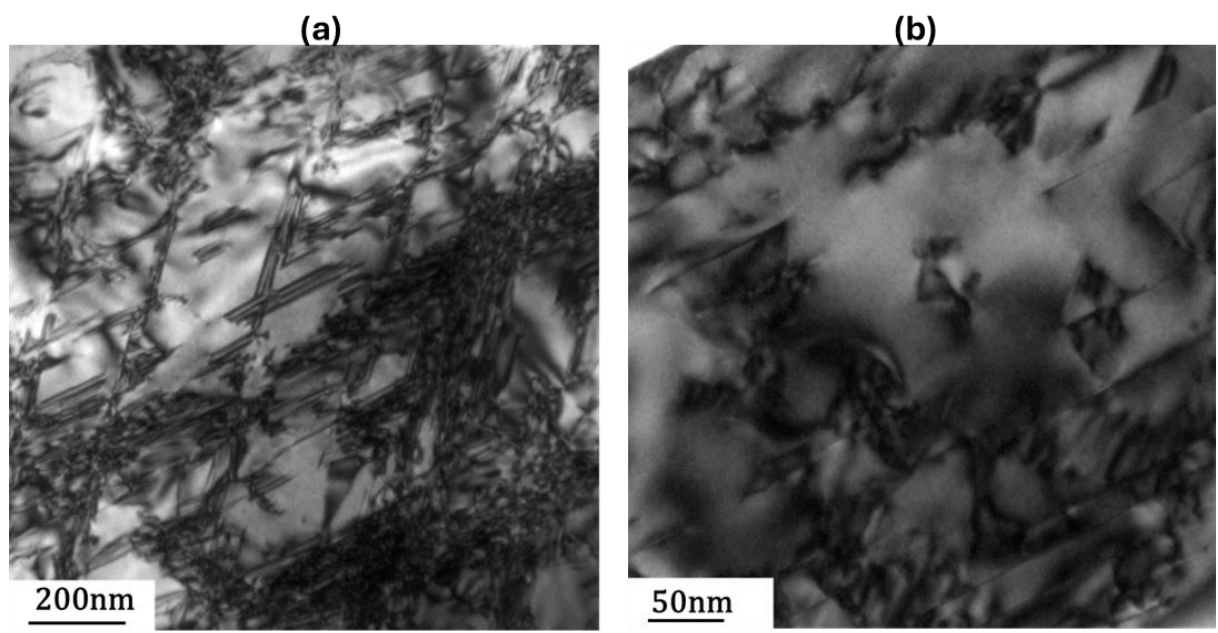

**Supplementary Fig. S8**: **a** BF-TEM micrograph taken along the [112] zone axis **b** BF-TEM image captured near to [110] zone axis showing several SFs

**S9: Bright-field TEM images of 25CR alloy showing stacking faults (SFs)**

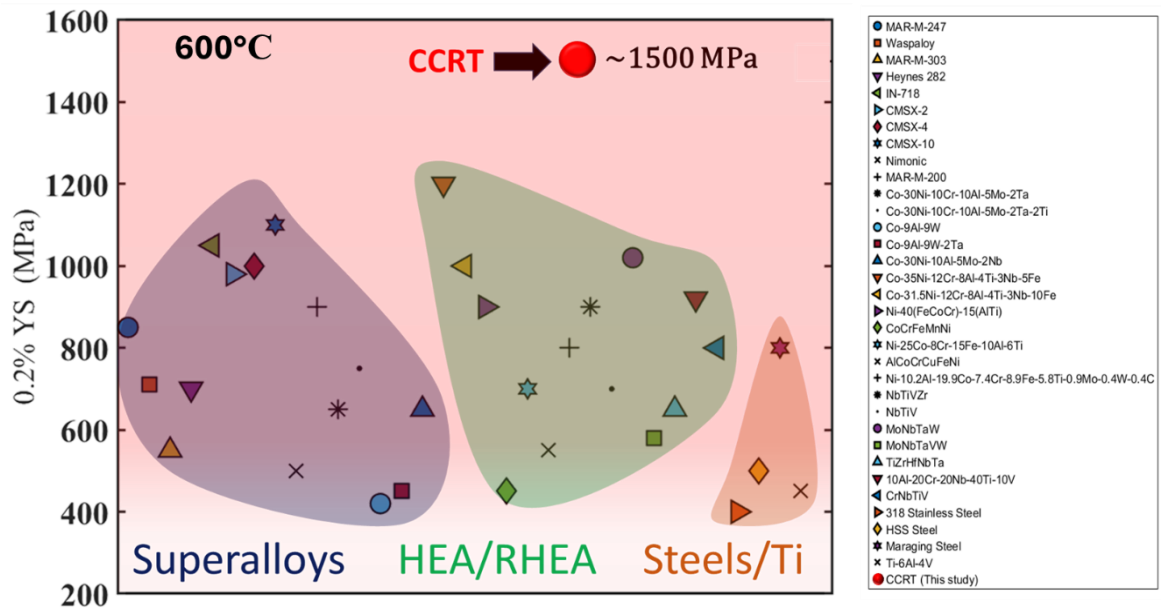

**Supplementary Fig. S9:** Comparison of 0.2% YS for CCRT alloy tested at 600°C with other structural and high temperature superalloys, multi/refractory, steels and Ti alloys.